\documentclass[aip,preprint]{revtex4-1}
\usepackage{graphicx}

\draft 

\begin{document}

\title{GPU acceleration of ab initio simulations of large-scale identical particles based on path integral molecular dynamics}

\author{Yunuo Xiong}
\email{xiongyunuo@hbpu.edu.cn}
\affiliation{Center for Fundamental Physics and School of Mathematics and Physics, Hubei Polytechnic University, Huangshi 435003, China}

\date{\today}

\begin{abstract}
Path integral Monte Carlo (PIMC) and path integral molecular dynamics (PIMD) provide the golden standard for the \textit{ab initio} simulations of identical particles. In this work, we achieved significant GPU acceleration based on PIMD, which is equivalent to PIMC in the \textit{ab initio} simulations, and developed an open-source PIMD code repository that does not rely on any other third party library. Numerical experiments show that for a system of 1600 interacting identical bosons in a harmonic trap, using a single GPU and a single CPU, it only takes two hours to achieve satisfactory simulation accuracy. With the increase of the number of identical particles, the advantage of GPU acceleration over CPU becomes more obvious, making it possible to simulate tens of thousands of identical particles from first principles using a single GPU. For example, for a system of 10000 non-interacting bosons, numerical experiments show that it takes 23 hours to obtain a simulation that is highly consistent with the exact results. Our study shows that GPU acceleration can lay a solid foundation for the wide application of PIMD simulations for extremely large-scale identical particle quantum systems with more than 10,000 particles. Numerical experiments show that a 24GB GPU can simulate up to 40000 identical particles from first principles, and the GPU acceleration leads to a roughly linear relationship between the computation time and the number of identical particles. In addition, we have also successfully implemented simulations for fictitious identical particle thermodynamics using GPU to overcome the Fermion sign problem, which makes it promising to efficiently and accurately simulate tens of thousands of fermions based on GPU.
\end{abstract}

\maketitle

\section{Introduction}

Path integral Monte Carlo (PIMC) and path integral molecular dynamics (PIMD) for the \textit{ab initio} simulations of identical particles provide the golden standard for simulating quantum systems \cite{CeperleyBook}. However, for large-scale quantum systems with thousands of identical particles, numerical simulations based on PIMD or PIMC generally require parallel computing on hundreds or even thousands of CPUs of a server cluster \cite{HirshImprove}; or even tens of thousands of CPUs of a supercomputer \cite{Dornheim2}. This situation seriously hinders the wider application of PIMC/PIMD in researchers' simulation of large-scale identical particles. For extremely large-scale quantum systems with tens of thousands of identical particles, even PIMC or PIMD simulation on supercomputers is highly lacking. Based on PIMC/PIMD, efficiently simulating systems with thousands or even tens of thousands of identical particles has become an urgent and highly challenging problem to be solved. Once this problem is solved, it will have unimaginable value for the \textit{ab initio} simulation of large-scale and extremely large-scale quantum systems, thus accelerating the development of quantum technology. Especially for the countless researchers who do not have the opportunity to use a large number of CPUs in supercomputers for parallel computing, the solution of this problem will stimulate the generation and application of a large number of innovative ideas.

It is well known that GPUs provide an alternative to CPUs for scientific computing \cite{Owens,Nickolls}, for example, GPU acceleration is indispensable in large language models. GPUs also have some applications in quantum physics, quantum chemistry, and materials science \cite{Maia,Esler,Shee,Andrade,Tamascelli,Lutsyshyn,Babich,Clark,Block,Giannozzi,Villalonga,Phillips,Le,Ufimtsev}. Considering the golden standard that PIMC/PIMD provides for \textit{ab initio} simulations, it is a natural question whether GPU can bring acceleration to PIMC/PIMD simulations. Unfortunately, due to the challenges of using GPU for PIMC/PIMD simulations, researches and applications on using GPU to simulate identical particle quantum systems with PIMC or PIMD have been highly lacking in the past two decades. In 2011, Quinn and Abarbanel \cite{Quinn} studied GPU acceleration for PIMC, but they did not explore the possibility of acceleration brought by quantum systems, and only demonstrated the Hodgkin-Huxley neuronal model, which is much simpler than identical particle quantum systems. Since GPUs are generally believed to play a role mainly in matrix operations, the mainstream view is that GPU may not be suitable for the \textit{ab initio} simulations of large-scale identical particle systems by PIMC/PIMD. This situation has led to the fact that the role of GPUs in quantum physics-related fields is far from comparable to that in artificial intelligence.

In this paper, we have successfully achieved GPU acceleration for large-scale and extremely large-scale identical particles based on PIMD, which is equivalent to PIMC in \textit{ab initio} simulations. For example, for 1600 weakly interacting bosons in a harmonic trap, our numerical experiments show that using a single GPU (24GB GeForce RTX4090) and a single CPU (Intel\textregistered Xeon\textregistered Gold 6226R 2.9G), it takes only two hours to obtain satisfactory simulation results in terms of energy (0.2\% error) and density distribution in a single simulation. For tens of thousands of bosons, GPU acceleration has even more significant value compared to massive CPU-based parallel computing based on PIMD. For example, the 24GB GPU used in our numerical experiments can accelerate the $\textit{ab initio}$ simulation of 40,000 identical bosons by 202 times.

Recently, fictitious identical particles \cite{XiongFSP,Xiong-xi} have been playing a significant role in simulating the thermodynamic properties of fermionic systems by overcoming the Fermion sign problem \cite{ceperley,troyer,Dornheim,Alex,WDM}, which has been experimentally verified and confirmed by the National Ignition Facility \cite{Dornheim3} to play a key role in important quantum systems such as inertial confinement fusion and red giants. Based on the $\xi$-extrapolation method, it has been demonstrated in the latest breakthrough \cite{Dornheim1,Dornheim2,Dornheim3,Dornheim4,Dornheim5} by Dornheim et al., to be a powerful tool for the \textit{ab initio} simulations of solid hydrogen and strongly compressed beryllium. Therefore, in this work, we also consider the GPU acceleration of PIMD simulations of fictitious identical particles in general, and find that it has the same GPU acceleration effect as identical bosons. This work lays a solid technical foundation for researchers to widely adopt the $\xi$-extrapolation method to study the thermodynamic properties of large-scale fermionic systems. Of course, the GPU acceleration is also useful for the \textit{ab initio} simulations of large-scale bosonic systems.

The organization of this paper is as follows.  In Sec. \ref{partitionS}, we briefly introduce the parameterized partition function for fictitious identical particles based on the path integral representation.
In Sec. \ref{GPU}, we present a GPU parallel computing scheme with implementation written in C and OpenCL for simulating fictitious identical particles using PIMD. In Sec. \ref{results}, we present some simulation results for identical particles and the effect of GPU acceleration. We compare the time costs of performing the same simulation on a single GPU and a single CPU, and find that for large-scale quantum systems, the acceleration effect of GPU is proportional to the number of particles. In Sec. \ref{summary}, we give a brief summary and discussion.

\section{Parameterized partition function for fictitious identical particles based on path integral representation}
\label{partitionS}

Recently, we have proposed fictitious identical particles \cite{XiongFSP,Xiong-xi} to overcome the Fermion sign problem, which opens up the possibility of efficiently and accurately simulating the thermodynamic properties of important fermionic systems \cite{Dornheim1,Dornheim2,Dornheim3,Dornheim4,Dornheim5} from first principles. Fictitious identical particles provide a unified mathematical framework for describing bosons, fermions, and distinguishable particles. Therefore, in this work, we will consider the GPU computation and acceleration effect of fictitious identical particle thermodynamics in general. The GPU acceleration presented here can be also directly applied to the \textit{ab initio} simulations of the thermodynamic properties of identical bosons. 

In fictitious identical particle thermodynamics \cite{XiongFSP,Xiong-xi}, we introduce a real parameter $\xi$ to describe the fictitious identical particles. $\xi=1$ represents bosons, $\xi=-1$ represents fermions. $\xi$ is a real number that can vary continuously. Fictitious identical particles with $\xi\neq \pm 1$ do not exist in the real world of elementary particles, but the thermodynamic properties of fictitious identical particles can be studied mathematically. 

We consider the following partition function:
\begin{equation}
Z(\beta)=Tr(e^{-\beta\hat H}).
\end{equation}
Here $\beta=1/k_BT$ with $k_B$ the Boltzmann constant and $T$ the temperature. The parametrized partition function for single-component fictitious identical particles with a real parameter $\xi$ can be written as \cite{XiongFSP,Xiong-xi,Dornheim1,Dornheim3,Dornheim4,Dornheim5}
\begin{equation}
Z(\xi,\beta)\sim\sum_{p\in S_N}\xi^{N_p}\int d\textbf{r}_1d\textbf{r}_2\cdots d\textbf{r}_N\left\langle p\{\textbf{r}\}|e^{-\beta\hat H}|\{\textbf{r}\}\right\rangle.
\label{Xipartition}
\end{equation}
$S_N$ represents the set of $N!$ permutation operations denoted by $p$. The factor $\xi^{N_p}$ is due to the exchange effect of fictitious identical particles. $\xi=+1$ for bosonic partition function, while $\xi=-1$  for fermionic partition function. We adopt the symbol convention $0^0=1$ here so that distinguishable particles ($\xi=0$) can also be included in the above expression. In addition, $\{\textbf{r}\}$ denotes $\{\textbf{r}_1,\cdots,\textbf{r}_N\}$. $N_p$ is a number defined to be the minimum number of times for which pairs of indices must be interchanged to recover the original order $\{\textbf{r}\}$ from $p\{\textbf{r}\}$. 

For the above parametrized partition function, we can decompose $e^{-\beta\hat H}$ as follows:
\begin{equation}
e^{-\beta\hat H}=e^{\Delta \beta\hat H}\cdots e^{\Delta \beta\hat H}.
\end{equation}
Here $\Delta\beta=\beta/P$, which means that there are a total of $P$ terms of $e^{\Delta \beta\hat H}$ multiplied together. After inserting the appropriate unit operators for momentum and coordinates,
based on the path integral formalism and generalizing the recursive method in Ref. \cite{HirshImprove}, the discretized partition function of Eq. (\ref{Xipartition}) is found to be:
\begin{equation}
Z(\xi,\beta)=\left(\frac{mP}{2\pi\hbar^2\beta}\right)^{PdN/2} \int e^{-\beta(V_\xi^{[1,N]}+\frac{1}{P}U)}d\mathbf R_1...d\mathbf R_N,
\label{partition}
\end{equation}
where $\mathbf R_i$ represents the collection of ring polymer coordinates $(\mathbf r_i^1,...,\mathbf r_i^P)$ corresponding to the $i$th particle. $P$ denotes the number of beads for a single particle. The system under consideration has $d$ spatial dimensions. $V_\xi^{[1,N]}$ considers the exchange effects of the fictitious identical particles with parameter $\xi$ by describing all the possible ring polymer configurations. $U$ is the interaction between different particles, which is given by
\begin{equation}
U = \sum_{l=1}^P V(\mathbf r_1^l,...,\mathbf r_N^l).
\end{equation}
Here $V$ denotes the interaction potential. The expression of $V_\xi^{[1,N]}$ is the key to consider the exchange effects of fictitious identical particles, which is given by the recursive relation shown in our previous paper \cite{Xiong-Hubbard}. 

For a given temperature $T$ (or $\beta=1/T$), in the bosonic sector ($\xi\geq 0$), $-\beta(V_\xi^{[1,N]}+\frac{1}{P}U)$ is a real number, so $ e^{-\beta(V_\xi^{[1,N]}+\frac{1}{P}U)}$  is a non-negative real function, which can be artificially assigned the meaning of a probability distribution. Since the integral factor in Eq. (\ref{partition}) is a function with $NPd$ variables, we often have to use Monte Carlo importance sampling when dealing with the above multiple integrals. For the importance sampling mentioned here, we have two ways: one is the usual Monte Carlo importance sampling, which leads to the development of PIMC \cite{Tuckerman,Fosdick,Jordan,barker,Morita,CeperleyRMP,Burov1,Burov1b}; the other is to use the molecular dynamics method, which is called PIMD. In this work, we will develop GPU computing based on PIMD \cite{HirshImprove,HirshbergFermi,Deuterium,Xiong2,Xiong5,Xiong6,Xiong7,Xiong4}, which is equivalent to PIMC in \textit{ab initio} simulations.

In this section, for the sake of simplicity, we take the case of spin polarization as an example to illustrate. It is straightforward to generalize spin polarization to the case of coexistence of different spins. The spin unpolarized case has been implemented in our open source code.

\section{GPU-based PIMD simulation scheme}
\label{GPU}

Generalizing the idea for identical bosons \cite{HirshImprove}, the expression of $V_\xi^{[1,N]}$ in Eq. (\ref{partition}) is the key to consider the exchange effects of fictitious identical particles, which is given by the following recursive relation \cite{Xiong-Hubbard}:
\begin{equation}
e^{-\beta V_\xi^{[1,N]}}=\frac{1}{N}\sum_{k=1}^N\xi^{k-1}e^{-\beta (E^{[N-k+1,N]}+V_\xi^{[1,N-k]})}.
\end{equation}
Here $E^{[N-k+1,N]}$ is defined as
\begin{equation}
E^{[N-k+1,N]}=\frac{1}{2}m\omega_P^2\sum_{l=N-k+1}^N\sum_{j=1}^P(\mathbf{r}_l^{j+1}-\mathbf{r}_l^j)^2,
\end{equation}
where $\mathbf{r}_l^{P+1}=\mathbf{r}_{l+1}^1$ unless $l=N$ where $\mathbf{r}_N^{P+1}=\mathbf{r}_{N-k+1}^1$. The harmonic strings connecting the ring polymers of different particles have the frequency $\omega_P=\frac{\sqrt P}{\beta\hbar}$.

The algorithm for PIMD simulation can be divided into several steps, each of which may be parallelized to yield a speedup over the sequential implementation. As a first step, we need to evaluate and store a number of quantities $E^{[N-k+1,N]}$ which are used for later calculations.
It is easy to see that for $N$ particles, the number of $E^{[N-k+1,N]}$ to be evaluated is $O(N^2)$, and calculating each one would take $O(NP)$ time, so a sequential evaluation of all $E^{[N-k+1,N]}$ takes $O(N^3P)$ time. It is also easy to see that the evaluation of each $E^{[N-k+1,N]}$ is independent, so we can launch as many threads as there are $E^{[N-k+1,N]}$ to evaluate each one simultaneously leading to a speedup over the sequential evaluation.
\par
Recently, a quadratic algorithm for identical bosons based on PIMD has been purposed by Feldman and  Hirshberg \cite{HirshImprove}, which eliminates the redundant calculations in $E^{[N-k+1,N]}$ by utilizing an iterative formula to calculate all $E^{[N-k+1,N]}$. A sequential implementation of this quadratic algorithm only takes $O(N^2+NP)$ time, and thus it is superior to the straightforward calculations. In this paper our primary focus is on this quadratic algorithm and its parallelization, so in the following we only discuss how to parallelize the quadratic algorithm, while the original $O(N^3P)$ algorithm is not discussed (a parallel implementation of the original algorithm is still available in our open source repository however \cite{XiongFSP}).
\par
The efficient method to evaluate $E^{[N-k+1,N]}$ works like this. First we have
\begin{equation}
E^{[v,v]}=E_{int}^{(v)}+\frac{1}{2}m\omega_P^2(\mathbf{r}_v^P-\mathbf{r}_v^1)^2,
\end{equation}
where
\begin{equation}
E_{int}^{(v)}=\frac{1}{2}m\omega_P^2\sum_{j=1}^{P-1}(\mathbf{r}_v^{j+1}-\mathbf{r}_v^j)^2.
\end{equation}
We can then evaluate the rest of $E^{[N-k+1,N]}$ iteratively as follows:
\begin{equation}
E^{[u,v]}=E^{[u+1,v]}-\frac{1}{2}m\omega_P^2(\mathbf{r}_v^P-\mathbf{r}_{u+1}^1)^2+\frac{1}{2}m\omega_P^2(\mathbf{r}_u^P-\mathbf{r}_{u+1}^1)^2+\frac{1}{2}m\omega_P^2(\mathbf{r}_v^P-\mathbf{r}_u^1)^2+E_{int}^{(u)}.
\end{equation}
A sequential evaluation of $E_{int}^{(v)}$ takes $O(NP)$ time and applying the iterative formula takes $O(N^2)$ time, so the overall complexity for the sequential program is $O(N^2+NP)$. 

Now we can consider how to parallelize the above calculations. First, it is straightforward to see that calculations for each $E_{int}^{(v)}$ (and also $E^{[v,v]}$, where $v=1,...,N$) are independent from each other, so we can launch $N$ threads to parallelize the calculations, no synchronization is required between different threads. Then we observe that evaluations of $E^{[u-1,u]}$ ($u=2,...,N$) depend on $E^{[v,v]}$ from the previous step, but they can be calculated independently from each other, so after we finished calculating $E^{[v,v]}$ we launch $N-1$ threads for $E^{[u-1,u]}$, no synchronization between threads is required in this case too. After that we can see $E^{[w-2,w]}$ ($w=3,...,N$) depend on $E^{[u-1,u]}$ and we can launch $N-2$ threads to parallelize the calculations. We continue all the way up to $E^{[1,N]}$, which terminates the iteration. Since minimal synchronization is required for the parallel calculations it is expected that the parallel scheme can provide an $N$ times speedup over the sequential implementation for moderate $N$.
\par
Once we have all the $E^{[N-k+1,N]}$, we define a set of potentials $V_\xi^{[u,N]}$ through the following recursive relation:
\begin{equation}
e^{-\beta V_\xi^{[u,N]}}=\sum_{l=u}^N\xi^{l-u}\frac{1}{l}e^{-\beta (E^{[u,l]}+V_\xi^{[l+1,N]})},
\end{equation}
where $V_\xi^{[N,N]}=0$ and $\xi$ is a real parameter characterizing fictitious identical particles. The potential $V_\xi^{[1,N]}$ results from the discretization of kinetic energy term in Feynman path integral for the partition function. We can see evaluating all potentials $V_\xi^{[u,N]}$ takes $O(N^2)$ time. To parallelize over this part of the algorithm we make use of the standard reduce add technique, which is a parallel routine to compute the sum over an array of elements $\sum_{i=1}^Na_i$, for moderate $N$ it effectively reduces the complexity of summation from $O(N)$ to $O(\log N)$, and so we simply use reduce add on the summation in the above formula for $V_\xi^{[u,N]}$, for each $u=1,...,N$. This way the computational complexity for the parallel program would no longer scales as $O(N^2)$ for moderate $N$. We also employ the same reduce add technique to compute the interaction energy $U$ which is a simple summation.
\par
In order to carry out molecular dynamics simulation we must also compute gradient of the potential for the force acting on each particle, which for $V_\xi^{[1,N]}$ is $-\nabla_{\mathbf{r}_l^j}V_\xi^{[1,N]}$. If we calculate the term $-\nabla_{\mathbf{r}_l^j}V_\xi^{[1,N]}$ for each particle directly based on the formula for $V_\xi^{[1,N]}$ it takes $O(N^2)$ time, and since there are $NP$ particles in total the final computational complexity is $O(N^3P)$. Fortunately, in the quadratic algorithm proposed by Feldman and Hirshberg \cite{HirshImprove}, an alternative and mathematically equivalent algorithm is used to compute all the $-\nabla_{\mathbf{r}_l^j}V_\xi^{[1,N]}$ efficiently, a sequential implementation of this part only takes $O(N^2+NP)$ time. 

The method works as follows. First we define a set of $N^2$ connection probabilities $Pr[G[\sigma](l)=l']$ for $\xi\geq 0$ with the following equations.
\par
For $l'\leq l$
\begin{equation}
Pr[G[\sigma](l)=l']=\xi^{l-l'}\frac{1}{l}\frac{1}{e^{-\beta V_\xi^{[1,N]}}}e^{-\beta(V_\xi^{[1,l'-1]}+E^{[l',l]}+V_\xi^{[l+1,N]})},
\end{equation}
while for $l'=l+1$
\begin{equation}
Pr[G[\sigma](l)=l+1]=1-\frac{1}{e^{-\beta V_\xi^{[1,N]}}}e^{-\beta (V_\xi^{[1,l]}+V_\xi^{[l+1,N]})},
\end{equation}
for other cases
\begin{equation}
Pr[G[\sigma](l)=l']=0.
\end{equation}

Each element of the $G$ matrix can be computed independently, and it is trivial to parallelize this part of the algorithm. Once we have all $Pr[G[\sigma](l)=l']$, we compute the gradients $-\nabla_{\mathbf{r}_l^j}V_\xi^{[1,N]}$ as follows. For exterior beads (particles with indices $j=1$ or $j=P$, for all $l=1,...,N$), we have
\begin{equation}
-\nabla_{\mathbf{r}_l^1}V_\xi^{[1,N]}=-\sum_{l'=1}^NPr[G[\sigma](l')=l]\cdot m\omega_P^2(2\mathbf{r}_l^1-\mathbf{r}_l^2-\mathbf{r}_{l'}^P),
\end{equation}
and
\begin{equation}
-\nabla_{\mathbf{r}_l^P}V_\xi^{[1,N]}=-\sum_{l'=1}^NPr[G[\sigma](l)=l']\cdot m\omega_P^2(2\mathbf{r}_l^P-\mathbf{r}_l^{P-1}-\mathbf{r}_{l'}^1).
\end{equation}
It is easy to parallelize over this part by doing the calculations for all $l$ simultaneously, an $N$ times speedup over the sequential routine is expected for moderate $N$. For interior beads (particles with indices $j=2,...,P-1$, for all $l=1,...,N$) we have
\begin{equation}
-\nabla_{\mathbf{r}_l^j}V_\xi^{[1,N]}=-m\omega_P^2(2\mathbf{r}_l^j-\mathbf{r}_l^{j+1}-\mathbf{r}_{l}^{j-1}).
\end{equation}
Again we can do the calculations for all $N\cdot(P-2)$ particles simultaneously, leading to a substantial speedup over the sequential implementation. 

Finally, we need to consider the interaction energy term:
\begin{equation}
U=\sum_{j=1}^P V(\mathbf{r}_1^j,...,\mathbf{r}_N^j).
\end{equation}
If pair interactions are included in $V(\mathbf{r}_1^j,...,\mathbf{r}_N^j)$, the total computational complexity for calculating $U$ would be $O(N^2P)$ (since $O(N^2)$ time is required for each individual $V$), which constitutes the most time consuming part of the simulation. Fortunately, $V(\mathbf{r}_1^j,...,\mathbf{r}_N^j)$ is just the classical interaction energy between $N$ particles with coordinates $(\mathbf{r}_1^j,...,\mathbf{r}_N^j)$, so any parallel implementation for classical molecular dynamics may be applied directly to calculate $U$ and its gradients. In our implementation we chose to calculate the contribution from each individual pair for the pair interaction simultaneously, which can achieve an $N$ times speedup over the sequential program for moderate $N$.

\begin{figure}[htbp]
\begin{center}
\includegraphics[scale=0.4]{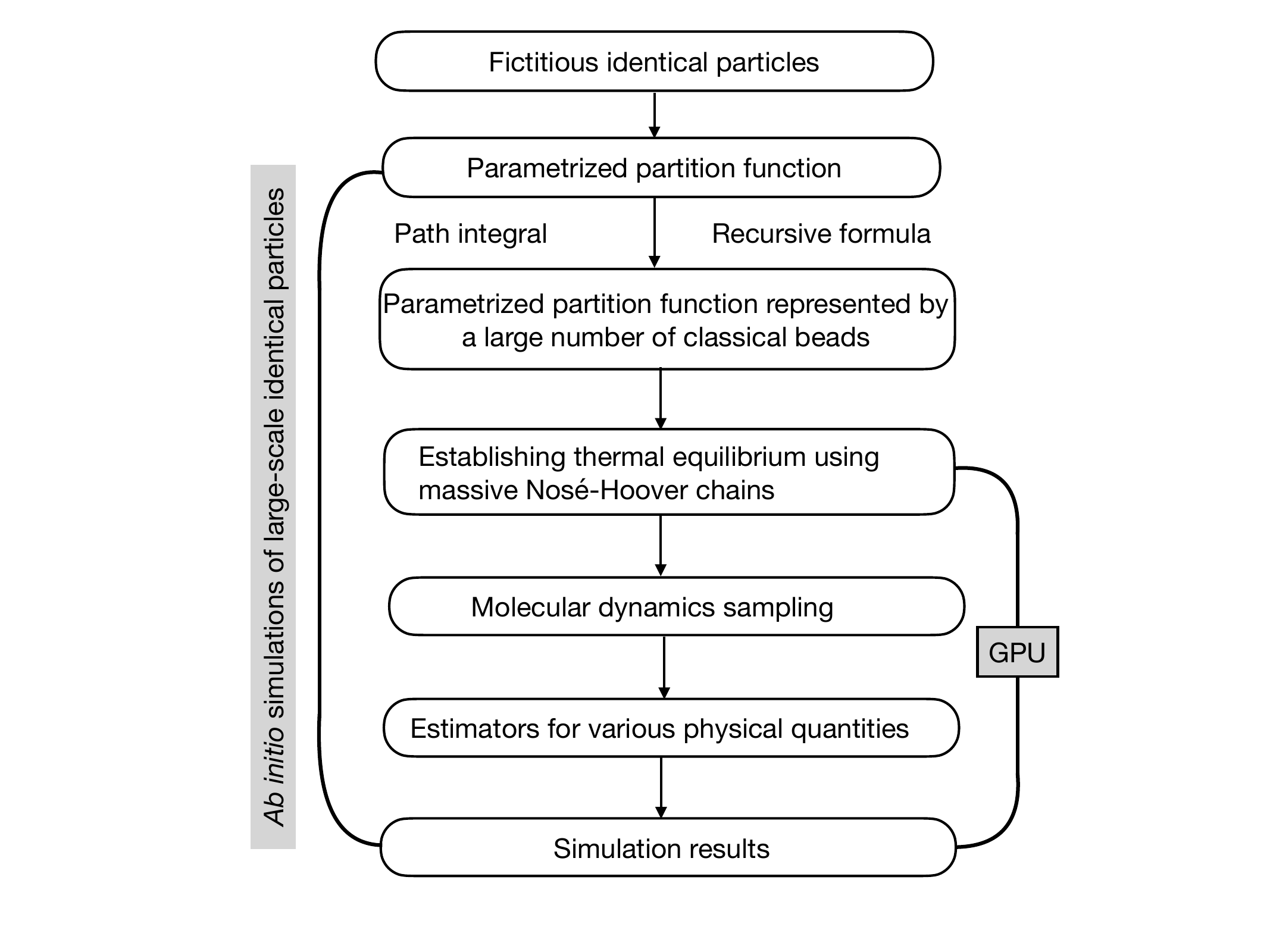}
\caption{Shown is the technical roadmap of GPU computing based on PIMD in this work.}
\label{structure}
\end{center}
\end{figure}

\par
In Fig. \ref{structure}, we summarize the technical roadmap of GPU computing based on PIMD in this work. This concludes our discussion of the efficient parallel algorithm for PIMD, all of the code implementing our scheme written in OpenCL \cite{OpenCL} are available in our open source repository. In the following we put our specific implementation to test and we observe that the computational time scales roughly linearly with number of particles $N$ for moderate $N$, instead of quadratically as with the sequential implementation, which verifies the expectation that our parallel algorithm can lead to an $N$ times speedup over the sequential algorithm. We also note that the memory usage of our program scales as $O(N^2)$, which may limit the scale of the system that can be simulated depending on the amount of available memory in a specific device.

\section{Results}
\label{results}

\subsection{Ideal Bose gas in a two-dimensional harmonic trap}
The Hamiltonian operator for a spinless ideal Bose gas in a two-dimensional harmonic trap is given by:
\begin{equation}
\hat{H}=-\frac{1}{2}\sum_{j=1}^N\Delta_j+\frac{1}{2}\sum_{j=1}^N\textbf{r}_j^2.
\end{equation}
We adopt the conventional convention $\hbar=k_B=m=\omega=1$ here. 

In our simulations, $P=72$ is used by default unless otherwise specified. The total MD (molecular dynamics) steps are expressed as $(a+b)$, where $a$ represents the MD steps taken in the process of establishing thermal equilibrium, and $b$ represents the MD steps taken after establishing thermal equilibrium to simulate thermodynamic properties by molecular dynamics sampling. This paper aims to demonstrate GPU acceleration and give technical solutions, so it does not seek to find the optimal ratio between $a$ and $b$. For specific quantum systems, we need to independently rely on experience or numerical experiments to determine the optimal ratio between $a$ and $b$. In this work, we simply choose the way of $a=b$.

\begin{figure}[htbp]
\begin{center}
\includegraphics[scale=0.3]{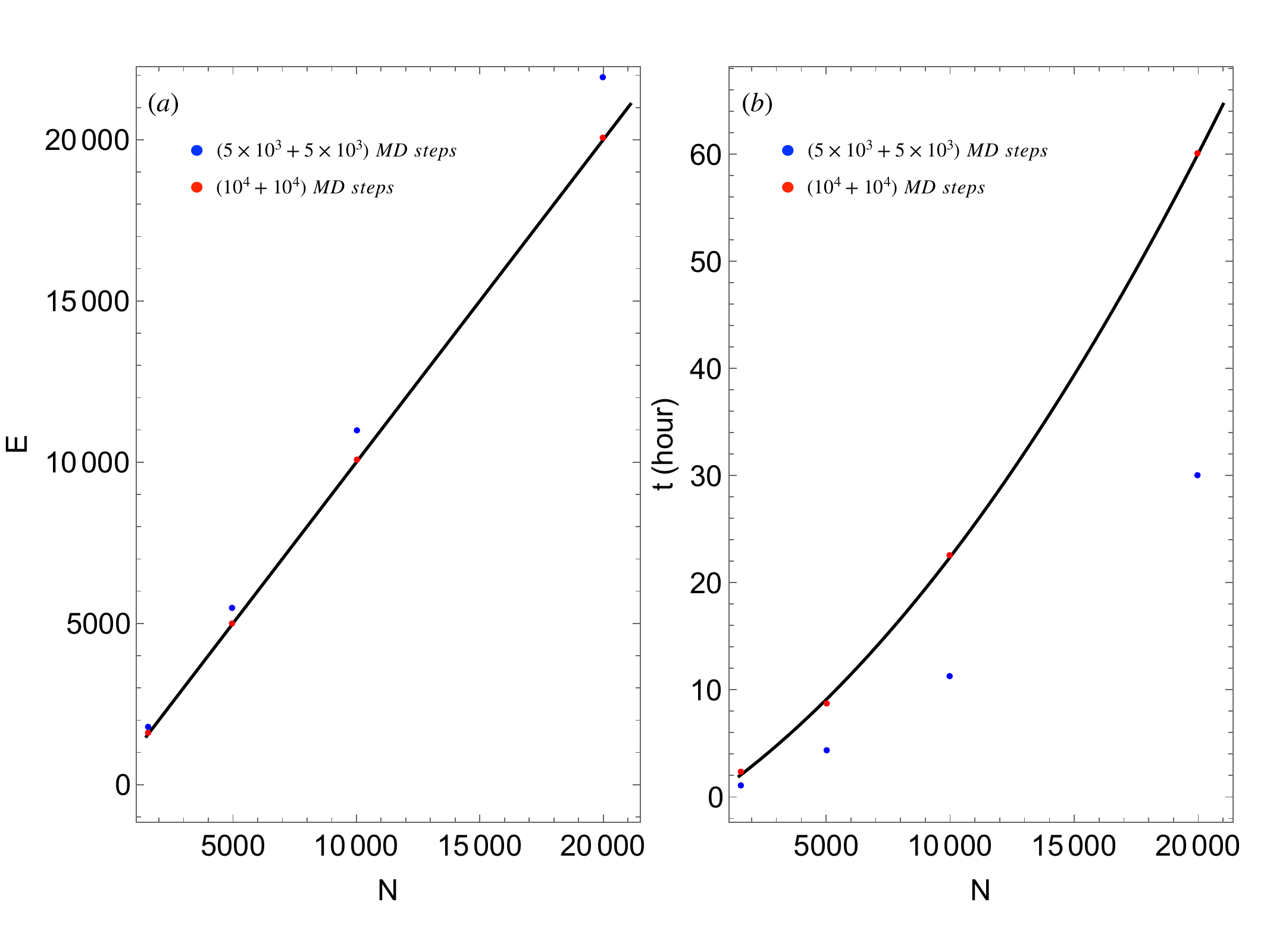}
\caption{Fig. (a) shows the energy simulation results based on GPU for different numbers of identical bosons. We calculated for $(10^4+10^4)$ MD steps and $(5\times 10^3+5\times 10^3)$ MD steps respectively. The red and blue dots are the energy results of PIMD simulation using GPU under different MD steps, and the black line represents the ground state energy $E=N$. The red and blue dots in Fig. (b) are the actual calculation time in hours of the MD steps corresponding to Fig. (a). The black solid line in Fig. (b) is a quadratic fit, and the fitting function is $t(N)=-0.565+0.00156N+7.36 \times10^{-8}N^2$, which is roughly linear.}
\label{idealenergy}
\end{center}
\end{figure}

In Fig. \ref{idealenergy}(a), we present the energies for different numbers of bosons at $\beta=6$ after a single simulation. For such a temperature $T=1/6$, we can well consider the system to be in the ground state due to $\hbar\omega>>k_BT$. The black line $E=N$ in the figure represents the analytical result of the ideal Bose gas in the ground state. The blue dots in the figure are the energy simulation results of $(5\times 10^3+5\times 10^3)$ MD steps, and the red dots are the energy simulation results of $(10^4+10^4)$ MD steps. The simulated energy given by the red dots is highly consistent with the exact ground state energy, with a deviation of about 0.3\%. There is an obvious deviation between the blue dots and the exact ground state energy, which is due to the insufficient MD steps and the failure to establish thermal equilibrium. We notice that for the red dots, once thermal equilibrium is established, the statistics of $10^4$ MD steps can give an accurate enough energy simulation result. The reason is that although the MD steps used for statistics are not large, due to the huge number of particles, it is equivalent to significantly improving the statistical data that can be used for energy simulation compared to the case of dozens of bosons. Therefore, once large-scale or extremely large-scale identical particles can be processed, the advantages of GPU in both computational efficiency and simulation accuracy can be greatly exploited due to the increase in statistical data.

In Fig. \ref{idealenergy}(b), we present the actual simulation time on the computer for different numbers of identical bosons. We use a GPU with 24GB of graphics memory (NVIDIA GeForce RTX4090) and a single CPU (Intel\textregistered Xeon\textregistered Gold 6226R 2.9G) to perform the calculation. The red dots represent the time consumed for $(10^4+10^4)$ MD steps, and the blue dots represent the time consumed for $(5\times 10^3+5\times 10^3)$ MD steps. We note that for a system of 1600 ideal bosons in a harmonic trap, it only takes a little over two hours to achieve satisfactory simulation accuracy. For a system of 1600 ideal bosons, Feldman and Hirshberg used a cluster of servers (each with two Intel Xeon Platinum 9242 CPU\textregistered 2.30GHz, 386GB RAM, and a total of 96 cores) for this case and took 9 days to complete $3\times 10^6$ MD steps. Our time for $3\times 10^6$ MD steps using the same $P=36$ as in the article \cite{HirshImprove} is 10 days, which is comparable. Of course, this comparison is not strictly meaningful, because the numerical experiment in the paper \cite{HirshImprove} used massive CPU parallel computing based on LAMMPS, while we used massive Nos\'e-Hoover chains \cite{Nose1,Nose2,Hoover,Martyna,Jang} to establish the thermal equilibrium for molecular dynamics and the code was completely independently written. This could also be one reason why we achieved satisfactory simulation results with much fewer than $3\times 10^6$ MD steps. It is worth noting that at present, people have not yet implemented GPU acceleration in PIMD based on LAMMPS, so we cannot compare with LAMMPS in terms of GPU acceleration.

From Fig. \ref{idealenergy}(b), we can see that the GPU computation time is roughly linear with the number of particles $N$, while Feldman and Hirshberg \cite{HirshImprove} summarized the computation time as $\sim N^{1.6}$ for the case of less than 1000 bosons. It is a pity that for the case of more particles, Feldman and Hirshberg did not give the relationship between the time required for parallel computing and the number of particles $N$ for the case of more than 1000 particles, due to the limitation of the communication speed between CPUs when using a large number of CPUs for parallel computing. In any case, based on the GPU acceleration here, our numerical experiments clearly show that a single GPU can be used to perform numerical simulations of more than 10,000 identical bosons, while the paper by Feldman and Hirshberg \cite{HirshImprove} did not give a simulation demonstration of such a extremely large-scale bosonic system.

\begin{figure}[htbp]
\begin{center}
\includegraphics[scale=0.4]{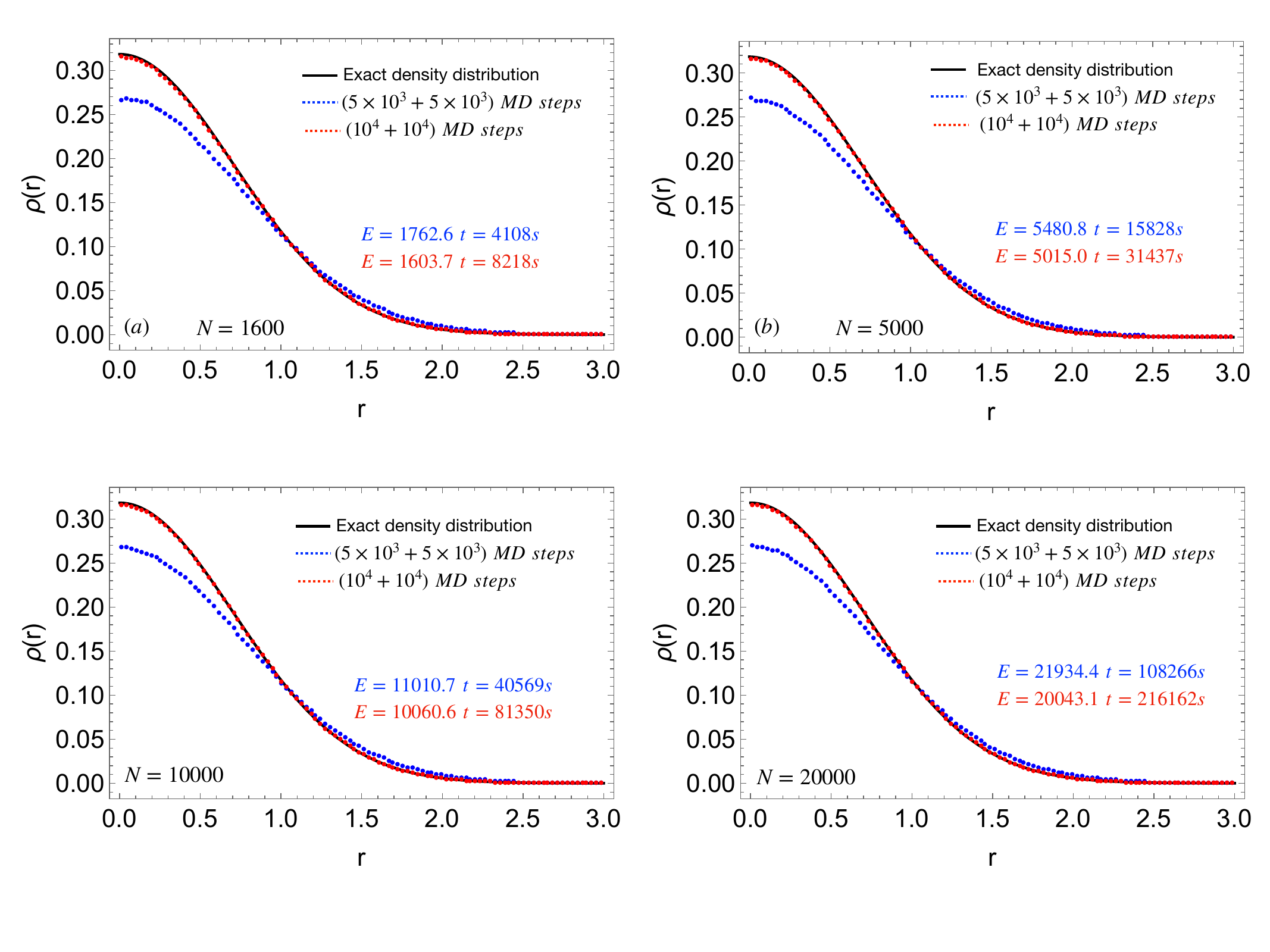}
\caption{
Shown are the GPU calculation results of PIMD for different numbers of identical bosons in a two-dimensional harmonic trap without interaction. The number of bosons corresponding to Figs. (a)-(d) are 1600, 5000, 10000, and 20000, respectively. In each figure, the black solid line is the analytical result of the normalized ground-state density distribution, and the blue dots are the simulation results for $(5\times 10^3+5\times 10^3)$ MD steps. The simulated energy and total time consumption are shown in blue font. The red dots in each figure are the simulation results for $(10^4+10^4)$ MD steps, and the simulated energy and total time consumption are shown in red font. For this bosonic system, the exact ground-state energy is $E=N$.}
\label{idealdensity}
\end{center}
\end{figure}

In Fig.~\ref{idealdensity}, we present the normalized density distribution and the analytical result of the ground-state density distribution (black solid line) $\rho(r)=\exp[-r^2]/\pi$, where $r=\sqrt{x^2+y^2}$. The red dots and blue dots in the figure are the simulation results of the density distribution for $(10^4+10^4)$ MD steps and $(5\times 10^3+5\times 10^3)$ MD steps, respectively. By comparing the red dots and the black solid line, we prove again that using GPUs for PIMD simulations can efficiently obtain highly accurate thermodynamic properties of large-scale identical Bose systems.

It is not surprising that PIMD can be used to accurately simulate the thermodynamic properties of identical bosons in large-scale Bose systems. In general, the relative fluctuation of a physical quantity $p$ is $1/\sqrt{M}$, where $M$ is the number of MD steps used in the sampling. It is worth noting that for energy and density distribution, when simulating a large number of identical particles, the number of particles $N$ can also contribute to the simulation accuracy. This is because when we obtain thermodynamic properties such as energy and density distribution through sampling, using a large number of particles can better suppress fluctuations. When the particle number $N$ is considered, the relative fluctuation of the physical quantity $p$ becomes $1/\sqrt{MN}$. This is why in the above simulation, only $10^{4}$
  MD steps of molecular dynamics sampling can obtain satisfactory results. This is good news for PIMD simulations of large-scale and extremely large-scale identical bosons, leading to a fairly satisfactory simulation accuracy for 1600 non-interacting bosons in a numerical experiment that took over two hours.

\subsection{Identical bosons with Gaussian interaction in a two-dimensional harmonic trap}

We now consider the following quantum system of identical bosons with Gaussian interaction:
\begin{equation}
\hat{H}=-\frac{1}{2}\sum_{j=1}^N\Delta_j+\frac{1}{2}\sum_{j=1}^N\textbf{r}_j^2+\frac{1}{2}\sum_{i\neq j}^{N} \frac{ g}{\pi s^2} e^{-|{\textbf r}_i-{\textbf r}_{j}|^2/w^2}.
\label{Gaussian}
\end{equation}

\begin{figure}[htbp]
\begin{center}
\includegraphics[scale=0.3]{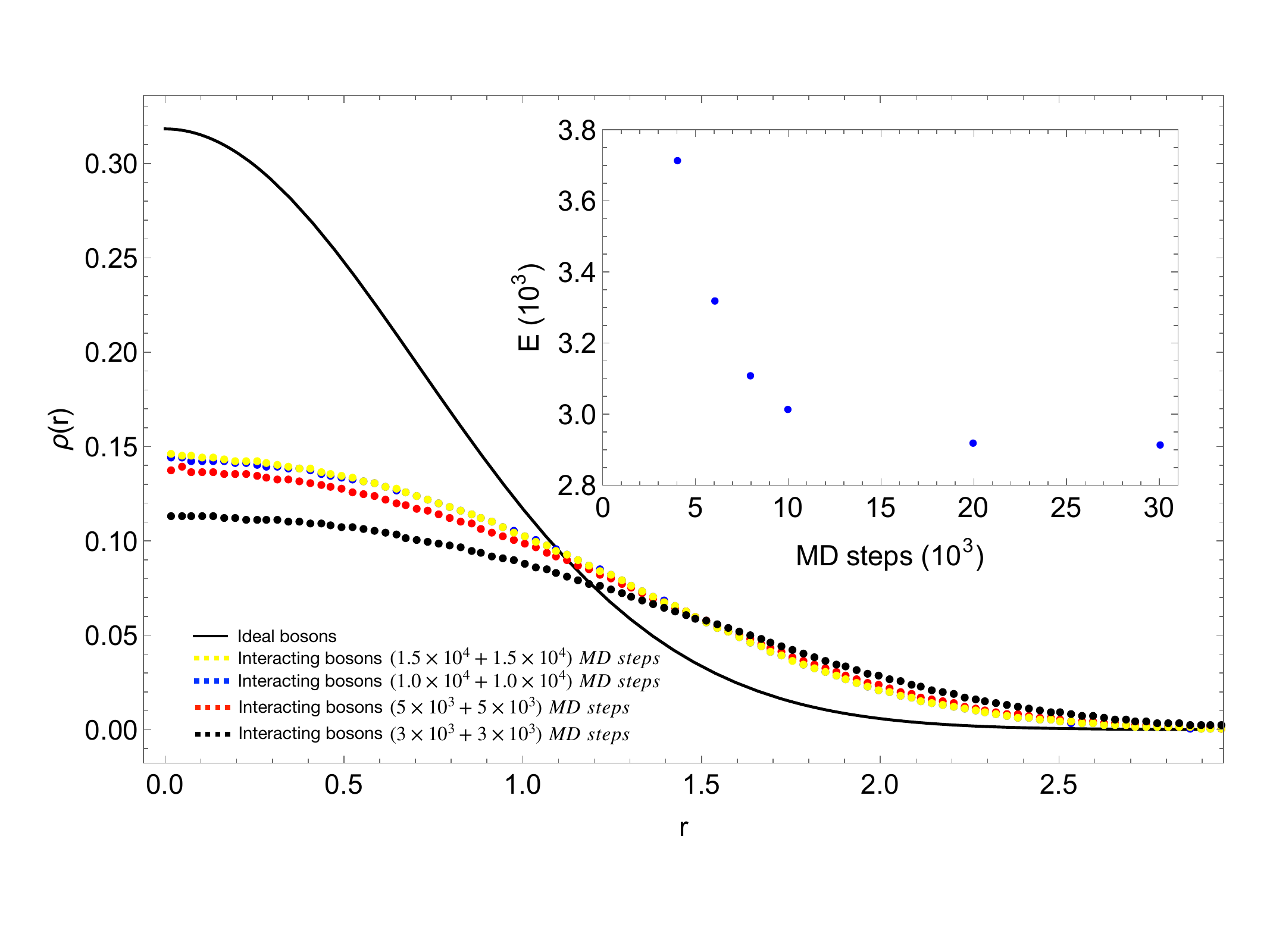}
\caption{For 1600 interacting bosons, the density distribution simulation results for different MD steps are shown in the figure after choosing $P=72$. The black solid line in the figure is the density distribution of the ideal Bose gas. In the inset, we show the relationship between the energy obtained from the \textit{ab initio} simulation and the total MD steps.
}
\label{1600intdensity}
\end{center}
\end{figure}

We take 1600 identical bosons as an example to carry out GPU calculations. In Fig.~\ref{1600intdensity}, we present the density distribution for the interaction parameters $(g=0.01,s=0.5)$ and the temperature $T=1/6$. The number of beads we choose is $P=72$. The black solid line in the figure is the ground-state density distribution of the ideal Bose gas, and the red dots, blue dots, yellow dots and black dots are the simulation results of the density distribution obtained after using different MD steps. We note that at $(10^4+10^4)$ MD steps, the convergence and sufficient statistical data can be guaranteed. Since the blue dots and yellow dots in the figure are highly coincident, many blue dots are covered by the yellow dots. In the inset of Fig.~\ref{1600intdensity}, we present the energy simulation results and the relationship between the total different MD steps. Since we consider the repulsive interaction here, the interaction leads to a wider density distribution compared to the ideal Bose gas. Similarly, the interaction also leads to an increase in the total energy compared to the ideal Bose gas.

\begin{figure}[htbp]
\begin{center}
\includegraphics[scale=0.3]{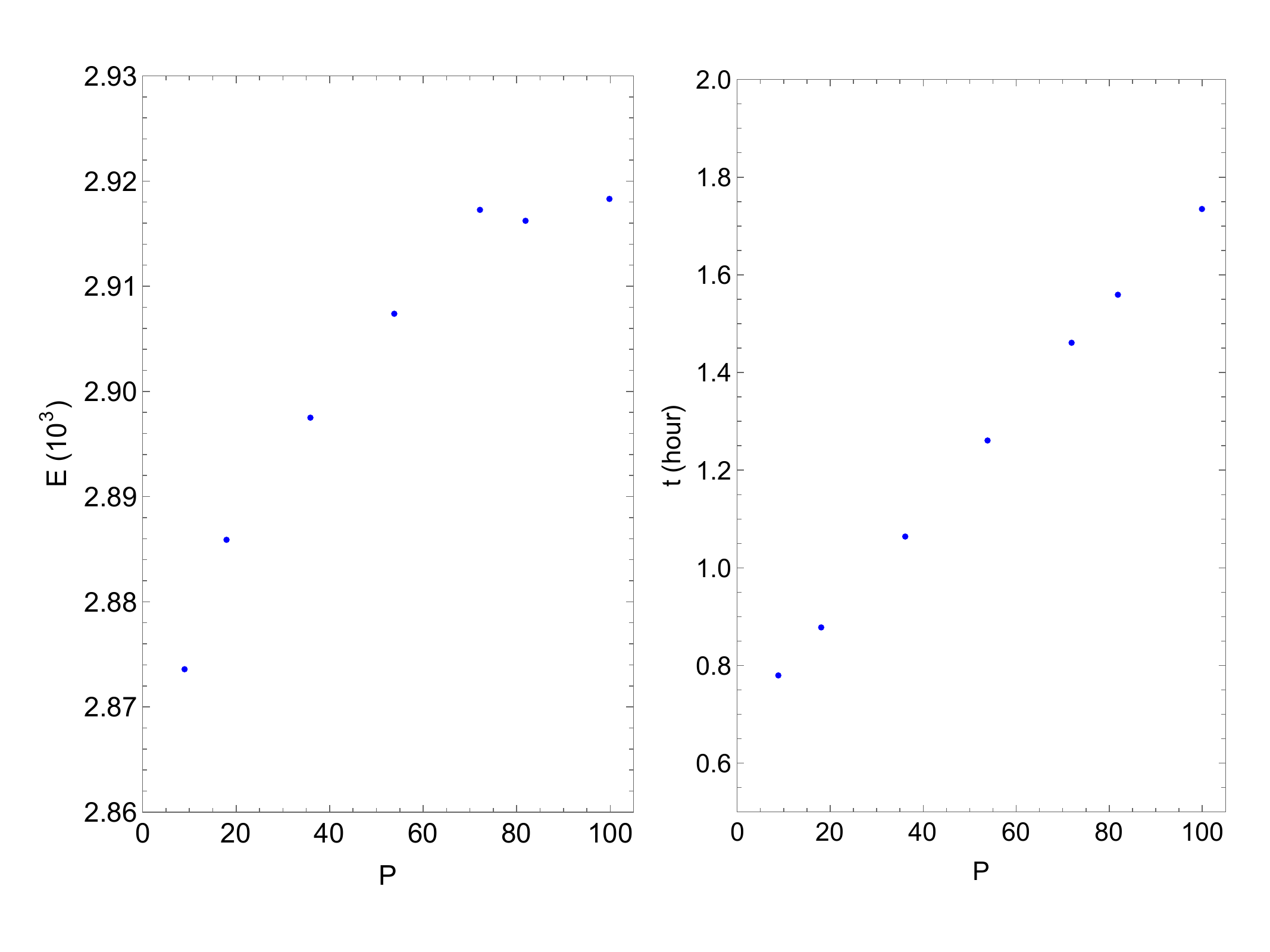}
\caption{
Fig. (a) shows the energy obtained from the PIMD simulation for different numbers of beads P for 1600 interacting bosons. Fig. (b) shows the actual calculation time based on GPU for different P under the same $(10^4+10^4)$ MD steps. The calculation time and $P$ satisfy a perfect linear relationship.}
\label{1600intenergyP}
\end{center}
\end{figure}

In Fig.~\ref{1600intenergyP}(a), the energy simulation results for different $P$ are presented at $(10^4+10^4)$ MD steps. We note that $P=72$ can guarantee convergence. In Fig.~\ref{1600intenergyP}(b), we present the relationship between the actual calculation time and $P$. We note that there is a linear relationship between the calculation time and $P$. The simulation of large-scale Bose systems can play a role in suppressing fluctuations, so all the results in this section are the energy and density distribution obtained from a single simulation. This paper demonstrates the convergence based on the example here, which does not mean that the selected MD steps and $P$ are applicable to all cases. In actual research, we need to verify the convergence of MD steps and $P$ for each case. In order to further improve the simulation accuracy, it is also an important method to do multiple simulations independently and then average them in some quantum systems.

\subsection{GPU simulation speedup}
\label{comparision}

\begin{figure}[htbp]
\begin{center}
\includegraphics[scale=0.7]{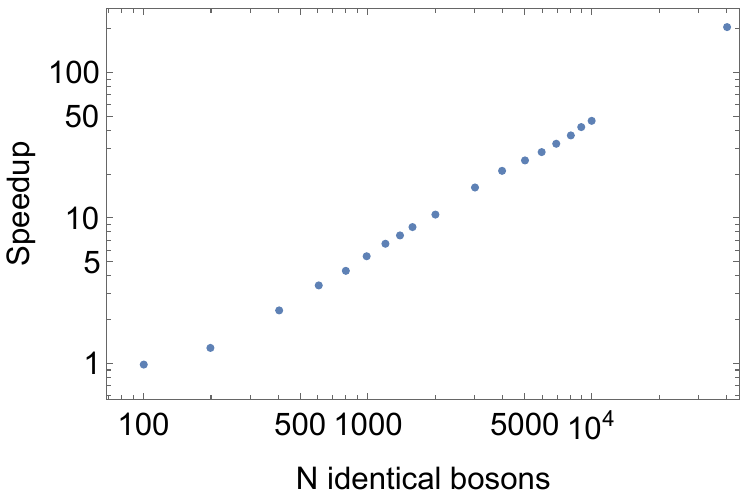}
\caption{
The figure shows the speedup of a single GPU compared to a single CPU for different numbers of bosons in the case of non-interacting bosons.}
\label{speedup}
\end{center}
\end{figure}

When measuring the acceleration brought by GPUs, a common metric is to compare it with the speed of a single CPU. In Fig.~\ref{speedup}, we present a comparison of numerical experiments performed on a single GPU (NVIDIA GeForce RTX4090) and a single CPU (Intel\textregistered Xeon\textregistered Gold 6226R 2.9G). The example compared here is the case of non-interacting particles in a two-dimensional harmonic trap.

As shown in Fig.~\ref{speedup}, the GPU acceleration is significant, and it can be observed that after more than 200 particles, the acceleration is roughly proportional to the number of particles. The reason for this is that the simulation time on a single CPU is roughly proportional to $N^2$, while after GPU acceleration, the simulation time is roughly proportional to $N$. This clearly demonstrates the significant advantages of GPUs in the \textit{ab initio} simulation of large-scale and extremely large-scale quantum systems.

\subsection{Fictitious identical particles with Gaussian Interaction in a two-dimensional harmonic trap}

In our previous work\cite{Xiong-Hubbard}, based on the new recursion formula for the partition function of identical bosons by Feldman and Hirshberg\cite{HirshImprove}, we have extended this new technique of recursion formula to fictitious identical particles to overcome the Fermion sign problem and simulate the thermodynamics of large-scale Fermi systems. For fictitious identical particles, we introduce a continuously varying real parameter $\xi$ to characterize the fictitious identical particles, $\xi=1$ for identical bosons, $\xi=0$ for distinguishable particles, and $\xi=-1$ for fermions. 

\begin{figure}[htbp]
\begin{center}
\includegraphics[scale=0.5]{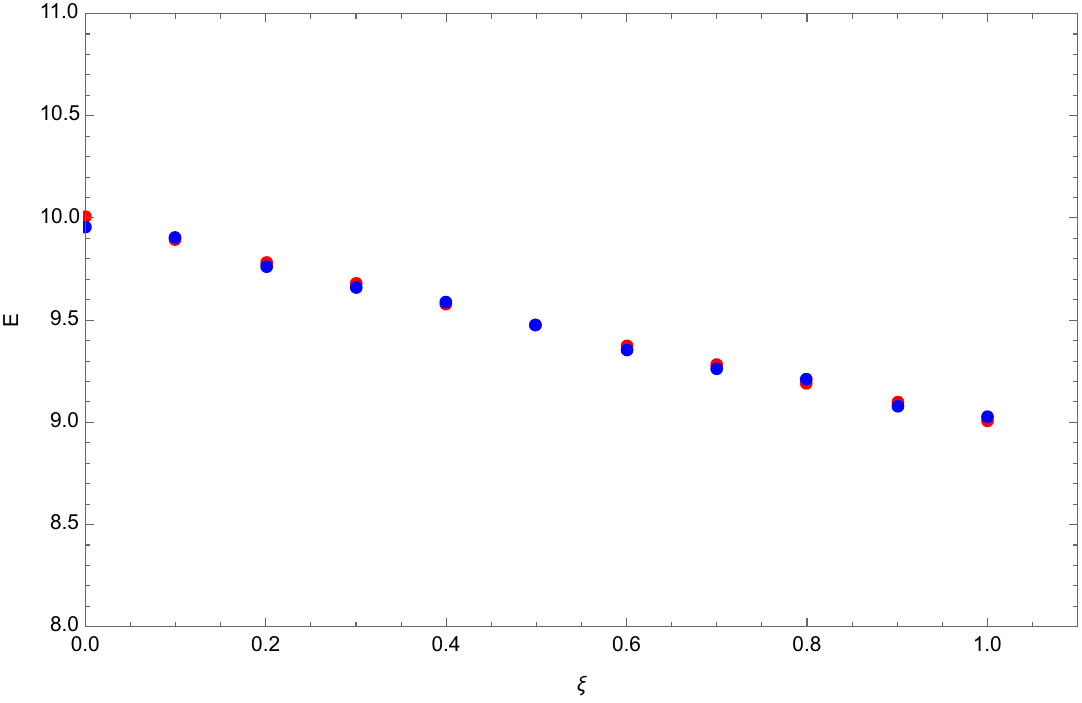}
\caption{
The blue dots in the figure are the energies of fictitious identical particles simulated by GPU in this work, and the red dots are the simulation results of our article\cite{XiongFSP} using CPU based on the partition function of the same fictitious identical particles.
}
\label{fparticles}
\end{center}
\end{figure}

For the case of $N=4,\beta=1,g=3,s=0.5$ with Gaussian interaction (Eq. (\ref{Gaussian})), in Fig.~\ref{fparticles} we compare the results calculated with GPU acceleration and those given in the article\cite{XiongFSP} without GPU acceleration. The two are highly consistent, which indicates that there is no error in the algorithm and code in our research with GPU acceleration. For the sake of comparison, as in the previous work\cite{XiongFSP}, here we averaged the energy of three independent simulations; in addition, $P=12$ and the total MD steps are $5\times 10^6$. We should note that for the case of very few particles, we need a large amount of data generated by $10^6$ MD steps to obtain results with sufficiently small fluctuations. In comparison, we find in this work that when the number of particles is in the thousands, as long as thermal equilibrium can be established, the data generated by $10^4$ MD steps can obtain accurate simulation results.

Since the topic of this paper is not to overcome the Fermion sign problem, we only use this simple example here to verify the correctness of the code and make it easy for researchers to independently repeat and verify the results here. Accelerating the thermodynamic calculation of fictitious identical particles with GPU and simulating the thermodynamic properties of Fermi systems is a future work worth focusing on.

\section{Summary and discussion}
\label{summary}

In summary, we have successfully implemented GPU acceleration of thermodynamics of fictitious identical particles in PIMD. With a single GPU, one can now have the capability that only server clusters or even supercomputers could have. In particular, for the \textit{ab initio} PIMD simulations of more than 10,000 identical particles, this work opens up a new technical implementation scheme. This work provides opportunities for countless researchers to conduct extensive \textit{ab initio} simulation studies of extremely large-scale quantum systems. Once parallel computing is performed with a large number of state-of-the-art GPUs, \textit{ab initio} exact numerical simulations of millions or even more identical-particle quantum systems will become a reality in the future. Due to the rapid development of fictitious identical particles in overcoming the Fermion sign problem to simulate large-scale fermionic systems, we have also implemented GPU simulation of the thermodynamics of fictitious identical particles here. In a series of groundbreaking works \cite{Dornheim1,Dornheim2,Dornheim3,Dornheim4,Dornheim5} by Dornheim et al., including experimental verification of the National Ignition Facility \cite{Dornheim3}, supercomputers were used to simulate the thermodynamics of fictitious identical particles when using the $\xi$-extrapolation method to simulate the thermodynamics of large-scale Fermi systems to a high-precision. This work opens up another way for the \textit{ab initio} simulations of Fermi systems of the same scale based on GPU. The code and data of this study are openly available in GitHub \cite{code}.

\begin{acknowledgments}
This work has received funding from Hubei Polytechnic University. I would like to thank Prof. Hongwei Xiong for conducting some numerical experiments and tests on the GPU computing code in this work.
\end{acknowledgments}


\begin{thebibliography}{10}

\bibitem{CeperleyBook} R. M. Martin, L. Reining, and D. M. Ceperley, \textit{Interacting Electrons: Theory and Computational Approaches} (Cambridge University Press, Cambridge, UK, 2016).


\bibitem{HirshImprove} Y. M. Y. Feldman and B. Hirshberg, Quadratic Scaling Bosonic Path Integral Molecular Dynamics, J. Chem. Phys. \textbf{159}, 154107 (2023).

\bibitem{Dornheim2} T. Dornheim, S. Schwalbe, Z. A.  Moldabekov, J. Vorberger, and P. Tolias, \textit{Ab initio} path integral Monte Carlo simulations of the uniform electron gas on large length scales, J. Phys. Chem. Lett. \textbf{15}, 1305 (2024).

\bibitem{Owens} J. D. Owens, M. Houston, D. Luebke, S. Green, J. E. Stone, J. C.  Phillips,  GPU computing, Proceedings of the IEEE \textbf{96}, 879-899 (2008).

\bibitem{Nickolls} J. Nickolls, W. J.  Dally, The GPU computing era, IEEE micro \textbf{30}, 56-69 (2010).


\bibitem{Maia} J. D. C. Maia, G. A. Urquiza Carvalho, C. P. Mangueira Jr, S. R. Santana, L. A. F. Cabral, G. B.  Rocha, GPU linear algebra libraries and GPGPU programming for accelerating MOPAC semiempirical quantum chemistry calculations, J. Chem. Theory Comput. \textbf{8}, 3072-3081 (2012).

\bibitem{Esler} K. Esler, J. Kim, D. Ceperley, L. Shulenburger, Accelerating quantum Monte Carlo simulations of real materials on GPU clusters, Comput. Sci. Eng. \textbf{14}, 40-51 (2010).

\bibitem{Shee} J. Shee, E. J. Arthur, S. Zhang, D. R. Reichman, R. A.  Friesner, Phaseless auxiliary-field quantum Monte Carlo on graphical processing units, J. Chem. Theory Comput. \textbf{14}, 4109-4121 (2018).

\bibitem{Andrade} X. Andrade, A.  Aspuru-Guzik, Real-space density functional theory on graphical processing units: Computational approach and comparison to Gaussian basis set methods, J. Chem. Theory Comput. \textbf{9}, 4360-4373 (2013).

\bibitem{Tamascelli} D. Tamascelli, F. S. Dambrosio, R. Conte, M.  Ceotto, Graphics processing units accelerated semiclassical initial value representation molecular dynamics, J. Chem. Phys.  \textbf{140}, 174109 (2014).

\bibitem{Lutsyshyn} Y. Lutsyshyn, Fast quantum Monte Carlo on a GPU, Comput. Phys. Commun. \textbf{187}, 162-174 (2015).

\bibitem{Babich} R. Babich, M. A. Clark, B. Joó, Parallelizing the QUDA library for multi-GPU calculations in lattice quantum chromodynamics, In SC'10: Proceedings of the 2010 ACM/IEEE International Conference for High Performance Computing, Networking, Storage and Analysis (pp. 1-11), IEEE (2010).

\bibitem{Clark} M. A. Clark, R. Babich, K. Barros, R. C. Brower, C.  Rebbi, Solving Lattice QCD systems of equations using mixed precision solvers on GPUs, Comput. Phys. Commun. \textbf{181}, 1517-1528 (2010).

\bibitem{Block} B. Block, O. Virnau, T.  Preis, Multi-GPU accelerated multi-spin Monte Carlo simulations of the 2D Ising model, Comput. Phys. Commun. \textbf{181}, 1549-1556 (2010).

\bibitem{Giannozzi} P. Giannozzi, O. Baseggio, P. Bonfà, D. Brunato, R. Car, I. Carnimeo, C. Cavazzoni, S. De Gironcoli, P. Delugas, F. Ferrari Ruffino, A. Ferretti, Quantum ESPRESSO toward the exascale, J. Chem. Phys. \textbf{152}, 154105 (2020).

\bibitem{Villalonga} B. Villalonga, D. Lyakh, S. Boixo, H. Neven, T. S. Humble, R. Biswas, E. G. Rieffel, A. Ho, and S. Mandrà,  Establishing the quantum supremacy frontier with a 281 pflop/s simulation, Quantum Sci. Technol. \textbf{5}, 034003 (2020).

\bibitem{Phillips} J. C. Phillips, D. J. Hardy, J. D. Maia, J. E. Stone, J. V. Ribeiro, R. C. Bernardi, R. Buch, G. Fiorin, J. Hénin, W. Jiang, and R. McGreevy, Scalable molecular dynamics on CPU and GPU architectures with NAMD, J. Chem. Phys. \textbf{153}, 044130 (2020).

\bibitem{Le} S. Le Grand, A. W. Götz, and R. C. Walker, SPFP: Speed without compromise—A mixed precision model for GPU accelerated molecular dynamics simulations, Comput. Phys. Commun. \textbf{184}, 374-380 (2013).

\bibitem{Ufimtsev} I. S. Ufimtsev, T. J. Martinez, Quantum chemistry on graphical processing units. 3. Analytical energy gradients, geometry optimization, and first principles molecular dynamics,  J. Chem. Theory Comput.  \textbf{5}, 2619-2628 (2009).

\bibitem{Quinn} J. C. Quinn, H. D.  Abarbanel, Data assimilation using a GPU accelerated path integral Monte Carlo approach, J. Comput. Phys. \textbf{230}, 8168-8178 (2011).



\bibitem{XiongFSP} Y. Xiong and H. Xiong, On the thermodynamic properties of fictitious identical particles and the application to fermion sign problem, J. Chem. Phys. \textbf{157}, 094112 (2022).  


\bibitem{Xiong-xi} Y. Xiong and H. Xiong, On the thermodynamics of fermions at any temperature based on parametrized partition function, Phys. Rev. E  \textbf{107}, 055308 (2023).


\bibitem{ceperley} D. M. Ceperley, Path Integral Monte Carlo Methods for Fermions, in \textit{Monte Carlo and Molecular Dynamics of Condensed Matter Systems}, edited by K. Binder and G. Ciccotti (Editrice Compositori, Bologna, Italy, 1996).

\bibitem{troyer} M.~Troyer and U. J.~Wiese, 
Computational Complexity and Fundamental Limitations to Fermionic Quantum Monte Carlo Simulations, 
{\text{Phys. Rev. Lett.} \textbf{94}, 170201} (2005).

\bibitem{Dornheim} T.~Dornheim, 
The Fermion sign problem in path integral Monte Carlo simulations: quantum dots, ultracold atoms, and warm dense matter, 
\text{Phys. Rev. E}~\textbf{100}, 023307 (2019).

\bibitem{Alex} A. Alexandru, G. Basar, P. F. Bedaque, and N. C. Warrington, 
Complex paths around the sign problem, 
Rev. Mod. Phys. \textbf{94}, 015006 (2022).

\bibitem{WDM} T. Dornheim, S. Groth, and M. Bonitz, The uniform electron gas at warm dense matter conditions, Phys. Rep. \textbf{744}, 1 (2018).

%
\bibitem{Dornheim3} T. Dornheim, T. D\"oppner, P. Tolias,
M. P. B\"ohme, L. B. Fletcher, Th. Gawne, F. R. Graziani, D. Kraus, M. J. MacDonald, Zh. A. Moldabekov,
S. Schwalbe, D. O. Gericke, and J. Vorberger, Unraveling electronic correlations in warm dense quantum plasmas, arXiv:2402.19113 (2024).

\bibitem{Dornheim1} T. Dornheim, P. Tolias, S. Groth, Z. A. Moldabekov, J. Vorberger, and B. Hirshberg, Fermionic physics from \textit{ab initio} path integral Monte Carlo simulations of fictitious identical particles,  J. Chem. Phys. \textbf{159}, 164113 (2023).

\bibitem{Dornheim4} T. Dornheim, S. Schwalbe, M. P. B\"ohme, Z. A. Moldabekov, J. Vorberger, and P. Tolias, Ab initio path integral Monte Carlo simulations of warm dense two-component systems without fixed nodes: structural properties, arXiv: 2403.01979 (2024).

\bibitem{Dornheim5} T. Dornheim, S. Schwalbe, P. Tolias, M. P. B\"ohme, Z. A. Moldabekov, J. Vorberger, Ab initio Density Response and Local Field Factor of Warm Dense Hydrogen, arXiv: 2403.08570 (2024).
%

\bibitem{Xiong-Hubbard} Y. Xiong, S. Liu and H. Xiong, Quadratic scaling path integral molecular dynamics for fictitious identical particles and its application to fermion systems, arXiv: 2401.00274 (2024).


\bibitem{Tuckerman} M. E.~Tuckerman, \textit{Statistical mechanics: theory and molecular simulation} (Oxford University, New York, 2010).

\bibitem{Fosdick} L. D. Fosdick and H. F. Jordan, Path-integral calculation of the two-particle Slater sum for He$^4$, Phys. Rev. \textbf{143}, 58–66 (1966).

\bibitem{Jordan} H. F. Jordan and L. D. Fosdick, Three-particle effects in the pair distribution function for He$^4$ gas, Phys. Rev. \textbf{171}, 128–149 (1968).


\bibitem{barker} J. A.~Barker, A quantum-statistical Monte Carlo method; path integrals with boundary conditions, J. Chem. Phys. \textbf{70}, 2914 (1979).

\bibitem{Morita} T. Morita, Solution of the Bloch Equation for Many-Particle Systems in Terms of the Path Integral, J. Phys. Soc. Japan. \textbf{35}, 980 (1973).

\bibitem{CeperleyRMP} D. M. Ceperley, Path integrals in the theory of condensed helium, Rev. Mod. Phys. \textbf{67}, 279 (1995).



\bibitem{Burov1} M. Boninsegni, N. V. Prokof’ev, and B. V. Svistunov, Worm Algorithm for Continuous-Space Path Integral Monte Carlo Simulations, \text{Phys. Rev. Lett.} ~\textbf{96}, 070601 (2006). 

\bibitem{Burov1b} M. Boninsegni, N. V. Prokof’ev, and B. V. Svistunov, Worm algorithm and diagrammatic Monte Carlo: A new approach to continuous-space path integral Monte Carlo simulations, Phys. Rev. E \textbf{74}, 036701 (2006).


\bibitem{HirshbergFermi} B. Hirshberg,  M. Invernizzi, and  M. Parrinello, 
Path integral molecular dynamics for fermions: Alleviating the sign problem with the Bogoliubov inequality, 
\text{J. Chem. Phys.} \textbf{152}, 171102 (2020).

 \bibitem{Deuterium}   C. W. Myung, B. Hirshberg, and M. Parrinello, Prediction of a supersolid phase in high-pressure deuterium, \text{Phys. Rev. Lett.} \textbf{128}, 045301 (2022).

\bibitem{Xiong2} Y. Xiong and  H. Xiong, 
Numerical calculation of Green's function and momentum distribution for spin-polarized fermions by path integral molecular dynamics, 
J. Chem. Phys. \textbf{156}, 204117 (2022).

\bibitem{Xiong5} Y. Xiong and H. Xiong, Path integral and winding number in singular magnetic field, Eur. Phys. J. Plus \textbf{137}, 550 (2022).

\bibitem{Xiong6} Y. Xiong and H. Xiong, Path integral molecular dynamics for anyons, bosons, and fermions, Phys. Rev. E \textbf{106}, 025309 (2022).

\bibitem{Xiong7} Y. Xiong and H. Xiong, Path integral molecular dynamics simulations for Green’s function in a system of identical bosons, J. Chem. Phys. \textbf{156}, 134112 (2022).

\bibitem{Xiong4} Y. Yu, S. Liu, H. Xiong, and Y. Xiong, Path integral molecular dynamics for thermodynamics and Green's function of ultracold spinor bosons, J. Chem. Phys. \textbf{157}, 064110 (2022).

\bibitem{OpenCL} A. Munshi, B. Gaster, T. G. Mattson, D. Ginsburg, \textit{OpenCL programming guide} (Addison-Wesley Professional, USA, 2011).

\bibitem{Nose1} S. Nos\'e, 
A molecular dynamics method for simulations in the canonical ensemble, 
\text{Mol. Phys.} \textbf{52}, 255 (1984).

\bibitem{Nose2} S. Nos\'e, 
A unified formulation of the constant temperature molecular dynamics methods, 
\text{J. Chem. Phys.} \textbf{81}, 511 (1984).

\bibitem{Hoover} W. G. Hoover, 
Canonical dynamics: Equilibrium phase-space distributions, 
\text{Phys. Rev. A} \textbf{31}, 1695 (1985).

\bibitem{Martyna} G. J. Martyna, M. L. Klein, and M. Tuckerman, 
Nos\'e-Hoover chains: The canonical ensemble via continuous dynamics, 
\text{J. Chem. Phys.} \textbf{97}, 2635 (1992).

\bibitem{Jang} S. Jang and G. A. Voth, 
Simple reversible molecular dynamics algorithms for Nos\'e-Hoover chain dynamics, 
\text{J. Chem. Phys.}~\textbf{107}, 9514 (1997).

\bibitem{code} https://github.com/xiongyunuo/MD-Lab


\end{thebibliography}
\end{document}